\documentclass[preprint,aps,psfig,nofootinbib,showpacs]{revtex4}
\usepackage{bm}
\usepackage{epsfig}
\bibstyle{apsrev}
\def\gev{{\rm \,Ge\kern-0.125em V}}

\def\half{{1\over2}}

\def\cG{{\cal G}}

\def\kslash{\mathbin{k\mkern-10mu\big/}}
\def\dslash{\mathbin{\partial\mkern-10mu\big/}}
\def\dis{}
\def\hatS{\hat S}
\def\hatS3{\hat S^3}
\def\hatSj{\hat S^j}
\def\hatbeta{\hat \beta}
\def\hattheta{\hat \theta}
\def\hatalpha{\hat \alpha}
\def\hatalphai{\hat \alpha^i}
\def\hatalpha3{\hat \alpha^3}
\def\hatthetaprime{\hat \theta^\prime}
\def\hatm{\hat m}
\def\hatp{\hat p}
\def\thetaprime{\theta^\prime}
\def\thetadprime{\theta^{\prime\prime}}
\def\doubleprime{{\prime\prime}}
\def\hi{\phantom{i}}
\def\Tr{{\rm {Tr}}}
\begin{document}

\begin{flushright}
\tt hep-ph/0412108v2
\end{flushright} 
\title{Studying Electroweak Baryogenesis using Evenisation and the
Wigner Formalism
}
\author{Jitesh Bhatt$^*$ and Raghavan Rangarajan$^\dagger$
}
\affiliation{Theoretical Physics Division, Physical Research Laboratory\\
Navrangpura, Ahmedabad 380 009, India}
\date{\today}
\begin{abstract}
We derive the kinetic equation for fermions and antifermions interacting
with a planar Higgs bubble wall during the electroweak phase transition
using the `evenisation' procedure and the Wigner formalism for a 
Lagrangian with the phase of the complex fermion mass rotated away.  
We obtain the energy, velocity and force for the particles in the presence of
the Higgs bubble wall.  Our results using both methods are in agreement.
This indicates the robustness of evenisation as a method to study quantum
corrections to the velocity and force for particles in the Higgs wall during
the electroweak phase transition.
We also derive the transport equations from the zeroth and first moment of
the kinetic equation.

\vspace{0.7cm}
{\tt
\noindent $^*$ jeet@prl.ernet.in\\
$^\dagger$ raghavan@prl.ernet.in\\}
\end{abstract}
\maketitle
\newpage
\setcounter{page}{1} 
\section{Introduction}

In models of electroweak baryogenesis, baryon asymmetry
is created during the electroweak
phase transition at temperatures of 
around 100 GeV.  Since this temperature corresponds 
to an energy scale that is experimentally accessible, there
has been considerable interest in  ascertaining the details of electroweak 
baryogenesis models,
in particular, 
the kinetic equation or Boltzmann equation for
the transport of particles through the Higgs bubble wall in a first order
electroweak phase transition.  

It is well known that, in the absence of collisions, the characteristics of the
Boltzmann equation
are the single particle trajectory equations, 
i.e. if $f_{}({\bf p},{\bf x},t)$ is a distribution 
function for a system of particles 
then $\frac{df_{}({\bf p},{\bf x},t)}{dt}=0$, 
where $\bf x$ and $\bf p$ satisfy single particle
equations of motion.  Expanding the total derivative
using the chain rule,
\begin{equation}
\partial_tf_{}\,+\frac{d {\bf x}}{dt}\partial_{\bf x} f_{}\,
+\frac{d{\bf p}}{dt}\partial_{{\bf p}}f_{}\,=\,0 \, ,
\label{BE}
\end{equation}
and $\frac{d {\bf x}}{dt}$ and $\frac{d {\bf p}}{dt}$ can be obtained
for the system
from the single particle equations of motion (Hamilton's equations).
However, in relativistic quantum theories it is non-trivial
to obtain the single particle equations of motion from the Hamiltonian
because of interference between particle and antiparticle states causing
Zitterbewegung.  But one may obtain
equations of motion that have a consistent
single particle interpretation
by making use of `evenised' operators \cite{grei}.
In Ref. \cite{br} we used evenisation to 
calculate 
particle trajectories to $O(\hbar)$
in the presence of the Higgs bubble
wall
during the electroweak phase
transition with a complex
mass term given by $m_R\bar\psi\psi +i m_I\bar\psi\gamma^5\psi
= |m| \bar\psi e^{i\theta \gamma^5}\psi$.
In this article we apply the procedure of evenisation to 
study the interaction of fermions
and antifermions with the Higgs bubble wall  
with the phase in the mass term rotated away and with an additional term
$\half \partial_\mu\theta \bar\psi\gamma^\mu\gamma^5\psi$ in the Lagrangian.  
We use this to obtain the energy relation $E({\bf p},{\bf x})$
and the kinetic equation.

Alternatively, in Ref. \cite{prok}, the energy 
relation, $\omega_{s \pm}({\bf p},{\bf x})$, and 
a kinetic equation 
in the absence of collisions
for particles
and antiparticles 
interacting with the Higgs bubble wall are obtained
from the equation of motion for the Wigner function $G^<(x,k)$,
which is the Wigner transform of the Wightman function 
$i\left< \bar{\psi}(x')\psi(x'') \right>$.
(Also see Refs. \cite{prok2,0406140}.) 
In this approach, particle and antiparticle distribution functions
are extracted from the Wigner function using a spectral decomposition that 
separates positive and negative energy states.
The particle and antiparticle distribution functions
are then found to
satisfy a collisionless quantum corrected kinetic equation
in the presence of the external force field provided by the Higgs
wall.  In Ref. \cite{prok} the Lagrangian was as in Ref. \cite{br},
i.e. with a complex pseudoscalar mass term, as mentioned above.
Here we apply the Wigner formalism to the rotated Lagrangian.  We obtain the
energy relation and kinetic equation.  These results agree with those
obtained using evenisation.  

The aim of this article is to establish the robustness of the 
application of the evenisation procedure to study quantum corrections
to particle trajectories.  Evenisation is a convenient technique that allows
us to reobtain classical expressions ($O(\hbar^0)$
for the velocity and force for a relativistic
quantum system of particles\cite{grei}.  
However, as far as we are aware, the extension of evenisation
techniques to obtain quantum corrections to the velocity and force for a system
of particles was first done in Ref. \cite{br} and shown to be consistent with
the Wigner formalism.  Below we work with a different Lagrangian and again find
consistent results between the two methods.

The energy relation and
kinetics (velocity and force) for a fermion in the bubble wall with the
real mass term were
first derived in Refs. \cite{jptcfm,9410282} using a WKB ansatz for the
components of the spinor wave function of the fermion.
Refs. \cite{jptcfm,9410282} considered states related with chirality
while below
we consider spin eigenstates.  However our results
are also compatible with the results 
in Refs. \cite{jptcfm,9410282} \footnote{Eq. (9) of Ref. \cite{9410282}
is missing an overall minus sign in the expression for the force in the
$z$ direction.}.

One may analyse the interaction of the particles with the background Higgs 
field with either the rotated or the unrotated Lagrangian.
The final baryon asymmetry is associated with the net axial current generated
in the wall and its outward diffusion 
away from the wall.  While
the Lagrangian is not invariant under the rotation, the axial current
$\langle \bar \psi \gamma^\mu\gamma^5\psi\rangle$ is invariant, 
and so the final asymmetry produced should be the same.

The energy relation obtained for the unrotated Lagrangian with a complex mass
is 
\begin{equation}
E\,=\,\sqrt{{p_z}^2
+|m|^2}-\frac{\hbar s}{2({p_z}^2+|m|^2)}|m|^2{\theta}^\prime
\end{equation}
and can not be easily recast as a mass shell condition.  The energy
relation for the rotated
Lagrangian with a real mass is
\begin{equation}
E\,=\,\sqrt{{p_z}^2+m^2}-{\hbar s\thetaprime\over2}
\end{equation}
and can be more easily restated as a mass shell condition
$(E+{\hbar s\thetaprime\over2})^2 -p_z^2=m^2$.  This is similar
to the mass shell condition for classical systems and may make it easier
to interpret the behaviour of the particles of the system.

\section{Kinetic equation using evenisation}
A first order electroweak phase transition proceeds
via the formation of Higgs bubbles.  As the bubbles expand they move through the
ambient sea of quarks, leptons and other particles.  
In this section we
first obtain a semi-classical expression
for the single particle
energy in terms of the evenised position and momentum.  
We then obtain expressions for the evenised velocity and force and
substitute these in Eq. (\ref{BE}) to obtain the kinetic equation.
A brief introduction to evenisation is given in Ref. \cite{br}.  Here we merely
present the necessary formulae for our analysis.

Any operator $\hat{A}$ can be split into an even part $\left[\hat{A}\right]$ 
and into
an odd part  $\left\{\hat{A}\right\}$. Even and odd parts of the operators are 
defined by
using the sign operator \cite{grei},
\begin{equation}
 \hat{\Lambda}\,=\,\frac{\hat{H}}{\sqrt {\hat H^2}} \, ,
\end{equation}
where $\hat H$ is the Hamiltonian.
The eigenvalues of the sign operator are $\pm1$, corresponding to particle and
antiparticle states.  Then
\begin{equation}
{\dis \left[\hat{A}\right]=\frac{1}{2}\left({\hat{A}}+\,
\hat\Lambda{\hat{A}}\hat\Lambda\right)}
\end{equation}
\begin{equation}
{\dis \left\{\hat{A}\right\}=\frac{1}{2}\left({\hat{A}}
-\,\hat\Lambda{\hat{A}}\hat\Lambda\right)}
=\frac{1}{2} [\hat A,\hat \Lambda]\hat\Lambda
\label{oddA}\end{equation}
The even part of the product of two operators $\hat{A} $ and ${\hat{B}}$ 
can be written as
\begin{equation}
\left[\hat{A}\hat{B}\right]\,=\,\left[\hat{A}\right]\left[\hat{B}\right]\,+
\left\{\hat{A}\right\}\left\{\hat{B}\right\} 
\label{prod}\end{equation}
We now apply the above to the system under consideration.

The Lagrangian describing the
interaction of particles with the bubble wall can be modeled by
\begin{equation}
{\cal L}=i\bar\psi \dslash\psi 
+\half \partial_\mu\theta \bar\psi\gamma^\mu\gamma^5\psi
-{m\over\hbar}\bar\psi\psi
\label{lagrangian}
\end{equation}
\noindent
This is a rotated form of the Lagrangian of Ref. \cite{br} with $m$ above
equivalent to $|m|$ in Ref. \cite{br}.
The Higgs bubble will be treated as a 
background field which provides in the bubble wall frame
a spatially varying real mass for the particles,
and a term associated with the axial current.
We shall consider the limit of large bubbles when the walls can be treated
as planar.
The corresponding Dirac equation 
\begin{equation}
(i \dslash 
+\half \partial_\mu\theta \gamma^\mu\gamma^5
-{m\over \hbar})\psi
=0
\label{Diraceq}
\end{equation}
can be rewritten in the form
\begin{equation}
i\hbar\frac{\partial \psi}{\partial t} = \hat H \psi \, ,
\end{equation}
with the Hamiltonian
$\hat{H}$ given by
\begin{eqnarray}
{\ \hat{H}}&=&\hat\alpha^i(\hat p^i+\frac{\hbar}{2} \hat 
{\partial^i\theta}\hat\gamma^5)
+ \hat\beta \hat m\\
&=&\hat\alpha^i\hat p^i+\frac{\hbar}{2} \hat {\partial^i\theta}\hat S^i
+ \hat\beta \hat m\, .
\end{eqnarray}
We have defined
the spin operator ${\bf\hat{S}}$ as ${\hat{\alpha}}\hat\gamma^5$, without
the standard $\hbar/2$ in the definition to be consistent with the notation in
Ref. \cite{prok}; the spin operator defined here is 
$\bf \hat \Sigma$ with Pauli matrices along the diagonal.  We have adopted the 
metric (1,-1,-1,-1). 
In this section $\hat{}$
designates an operator.
We now use the method of evenisation 
and evenise $\hat{H}^2$.  We discuss later why we 
evenise $\hat{H}^2$ rather than $\hat{H}$.

We first write after some algebra,
\begin{equation}
\hat{H}^2\,=\,{\bf{\hat{p}}}^2\,+\, \hat m^2\,+
\frac{\hbar^2}4(\hat {\partial^i\theta}\hat S^i)^2+
\hat\alpha^i\hat{\beta}\left[\hat p^i, \hat m\right]
\,+\hbar \hat{\partial^i\theta} \hat m \hat S^i\hat\beta
+{\hbar\over 2} \hatalphai\hatSj[\hat p^i,\hat{\partial^j\theta}]
+\hbar 
\hat{\partial^j\theta}\hatSj\hatalphai\hat p^i
\end{equation}
\noindent
We work in the wall frame and
take the wall to be planar in the $x-y$ plane and so 
$\hat m$ is a function of $\hat z$ only.
Then, keeping only terms to $O(\hbar)$, the Hamiltonian can be written as
\begin{equation}
\hat{H}^2\,=\,{\bf \hat{p}}^2\,+\, \hat m^2\,
-\hat S^3
\hat\gamma^0\hat\gamma^5(-i\hbar\hat{\partial_z m})\,
-\hbar\hatthetaprime  \hatm \hatS3\hatbeta
-\hbar\hatthetaprime\hatp^i \hatS3\hatalphai
 \, ,
\label{Hsquared}\end{equation}
where $z\equiv x^3$, $\partial_z\equiv\partial/\partial \hat x^3$
and $\hatthetaprime=\hat {\partial_z \theta}=-\hat{\partial^3\theta}$.
We chose to express $\hat\alpha^3\hat\beta$ 
in terms of $\hat S^3$
as this simplifies the evenisation of the corresponding term below.

We now evenise $[\hat H^2]$ to order $\hbar$.  
The spin operator in general does not commute with the Hamiltonian.
In order to simplify the problem, as in Ref.\cite{prok}, we assume
that we are in an inertial reference frame where  $x$ and $y$ 
components of the momentum
are zero and consequently $\hat \alpha^1$ and $\hat \alpha^2$ will be absent from
the Hamiltonian.  This will allow us to set $\{S^3\}$ to 0 later.
As mentioned earlier, we wish to express $[\hat H^2]$ in terms of
$[\hat p_z]$ and $[\hat z]$ (where $p_z\equiv p^3$).  
The odd part of $\hat p_z$ will be proportional
to $[\hat\Lambda,\hat p_z]$ (see Eq. (\ref{oddA}) above) and so will be of order
$\hbar$.  Therefore, using Eq. (\ref{prod}), we can approximate $[\hat p_z^2]$
by $[\hat p_z]^2$.  Similarly we approximate $[m(\hat z)^2]$ as
$[m(\hat z)]^2$.  Furthermore, since $\{\hat z\}$ is $O(\hbar)$,
$[m(\hat z)]=m([\hat z])+O(\hbar^2)$ as any dependence of the even operator
$[m(\hat z)]$ on the odd operator $\{\hat z\}$ can only appear as
$\{\hat z\}^2$.  (Expand $m(z)$ in a Taylor series about $z=0$ and then 
replace $z$ by $\hat z=[\hat z]+\{\hat z\}$.)
The last three terms on the right hand side of Eq. (\ref{Hsquared}) are 
$O(\hbar)$
and therfore we will use the sign operator defined upto $O(\hbar^0)$ to
evenise these terms.
We
define the zeroth order Hamiltonian and energy as
\begin{eqnarray}
{\ \hat H_0}&=
&\hat\alpha^3\hat p^3
+ \hat\beta \hat m\\
E_0&=&{\dis \sqrt{{{p_z}}^2\,+\,m^2}} \,
\label{energy0}\end{eqnarray}
where ${p_z}^2$ and $m^2$ are real numbers and are expectation values
of the corresponding operators in an eigenstate of definite energy and spin.
\noindent
With this we define 
the sign operator upto $O\left(\hbar^0\right)$ as follows
\begin{equation}
{\dis \hat{\Lambda}_0\,=\,\frac{\hat{H_0}}{\sqrt{{{p_z}}^2\,+\,m^2 }}}
\end{equation}
Note that $( \hat{\Lambda}_0)^2= 1+O(\hbar) $ which we shall use in the
derivation of evenised operators in the Appendix.
We obtain
\begin{equation}
[\hat \gamma^0\hat \gamma^5]\,=\,O(\hbar)
\end{equation}
\begin{equation}
[\hat\beta]\,=\,\frac{ \hat m}{E_0}\hat\Lambda_0\,+\,O(\hbar) \, .
\end{equation}
\begin{equation}
[\hat\alpha^3]\,=\,\frac{[\hat p_z]}{E_0}\hat\Lambda_0\,+\,O(\hbar) \, .
\end{equation}

Now
\begin{equation}
\,[\hat S^3]\,=\,[\hat{\alpha^3}\hat\gamma^5]\,=\,\hat S^3 \, ,
\end{equation}
i.e., $\{\hat S^3\}=0$ as $\hat S^3$ commutes with the Hamiltonian
(which is true only in the chosen inertial frame).  Then, keeping in mind
that the odd parts of $\hat m, \partial_z \hat m, \hat \thetaprime$ and
$\hat p^3$ that appear in the last three terms of $\hat H^2$ are $O(\hbar)$,
we get,
to $O(\hbar)$,
\begin{equation}
[\hat{H}^2]\,=\,[\hat{p_z}]^2\,+
\,  \hat m^2 \,
-\hbar\hatthetaprime{ \hat m^2\over E_0}\hatS3
 \hat\Lambda_0\,
-\hbar\hatthetaprime{[\hat p_z]^2\over E_0}\hatS3
 \hat\Lambda_0\, .
\end{equation}
where $\hat m$ and $\hatthetaprime$ are now functions of $[\hat z]$.
Replacing evenised operators $[\hat z]$ and $[\hat p_z]$ by 
real numbers representing their expectation values in
a state of definite energy and spin,
$|E,s\rangle$ for particles and $|-E,-s\rangle$ for antiparticles
\footnote
{In the earlier version of this paper, we had identified antiparticles of spin
$s$ with negative energy particle solutions of spin $s$, rather than
$-s$ (see Sec. 7.1 of Ref. \cite{grei}).
Hence the expressions for energy and force were different for particles
and antiparticles of the same spin.  
Our current results reflect the $P$ and $CP$ violation and $C$ conservation
properties of the Lagrangian in Eq. (\ref{lagrangian}).\label{negspin}}
,
we deduce the corresponding expression for the energy to be
\begin{equation}
E^2\,=\,{p_z}^2+m^2-\hbar s\thetaprime E_0 ,
\end{equation}
where we have also replaced $\hat S^3$ by its eigenvalue 
$\pm s$ and 
$\Lambda_0$ by $\pm1$ for particle/antiparticle states.  
\noindent
The energy relation for particles and antiparticles respectively
can then be written upto order $\hbar$ as
\begin{equation}
E\,=\,\sqrt{{p_z}^2+m^2}-{\hbar s\thetaprime\over2}
\label{energy}\end{equation}
Note that if
we had wished to obtain the energy relation by evenising
$\hat{H}$ instead of evenizing $\hat{H}^2$ we would have faced a problem as 
we would have needed 
$\hat \Lambda$ to $O(\hbar)$ which itself
requires $E$ to $O(\hbar)$.  
However, 
since $\{\hat H\}=0$, $[\hat H]=[\hat H^2]^{1/2}$.
We are now able to define the sign operator upto $O(\hbar)$ as 
$\hat\Lambda=\hat H/E$ and can use the same to obtain the evenised velocity
and force to $O(\hbar)$ for the kinetic equation.

We now obtain the kinetic equation to $O(\hbar)$.
Using the chain rule for partial differentiation
the kinetic equation is written as
\begin{equation}
\partial_tf_{s\pm}\,+\frac{dz}{dt}\partial_z f_{s\pm}\,
+\frac{dp_z}{dt}\partial_{p_z}f_{s\pm}\,=\,0 \, ,
\label{boltzeqn1}
\end{equation}
where $f_{s\pm}$ are the particle and antiparticle distribution functions.
We will associate $dz/dt$ and $d p_z/dt$ with the expectation values of 
$[d\hat z/dt]$ and $[d\hat p_z/dt]$ in states of definite energy and spin,
as before.
Implicitly we are assuming here that the
form of the quantum Boltzmann equation is the same as that of the
classical Boltzmann equation and that the quantum corrections are contained
only in the coefficients of the equation, namely, in the expressions for the
velocity and the force.  
The evenised expressions for the velocity and force to $O(\hbar)$
are derived in
the Appendix as
\begin{eqnarray}
[{d \hat z/dt}]&=&[\,-{i\over\hbar}[\hat z,\hat H]\,]\cr
&=&\left({ [\hat p_z] \over E} -{\hbar\hatthetaprime[\hat p_z]
\hatS3\hat\Lambda_0\over2E_0^2}
\right )\hat\Lambda
	      \end{eqnarray}
	      \begin{eqnarray}
[{d \hat p_z/ dt}]&=&[\,-{i\over\hbar}[\hat p_z,\hat H]\,]\cr
&=&
\left( -{ \hat m^{2\prime}\over 2E }
+{\hbar \hat m^{2\prime}\hat\theta'\hat\Lambda_0\hat S^3  
\over 4E_0^2}
+{\hbar\hat\theta^\doubleprime\hatS3\hat\Lambda_0\over2}
\right)\hat\Lambda\, .
\end{eqnarray}
Substituting the eigenvalues of the above
operators in the kinetic equation
\footnote{
The overall minus sign in the expectation value for the velocity and force
for antiparticles can be absorbed by taking the expectation value of the 
momentum in antiparticle states to be $-p_z$, as
for systems with constant mass, where the energy eignenstates are also
momentum eigenstates.
},
we get
\begin{equation}
\partial_tf_{s\pm}\,+{p_z\over E}
\left(1-{\hbar \thetaprime s\over 2E_0}\right)
\partial_z f_{s\pm}\,
+\left[-{  m^{2'}\over 2E} 
\left(1-{\hbar\thetaprime s\over 2 E_0}\right)
		+ {\hbar s \theta^{\prime\prime}\over 2  }
		\right ]
\partial_{p_z}f_{s\pm}\,=\,0 \, .
\label{boltzeqn2}
\end{equation}
Simplifying the above equation by using Eq. (\ref{energy}) to 
express $E$ as $E_0[1-\hbar\thetaprime s/(2E_0)]$ we get
\begin{equation}
\partial_tf_{s\pm}\,+{p_z\over E_0}
\partial_z f_{s\pm}\,
+\left(-{  m^{2'}\over 2E_0} 
		+ {\hbar s \theta^{\prime\prime}\over 2  }
		\right )
\partial_{p_z}f_{s\pm}\,=\,0 \, ,
\label{boltzeqn3}
\end{equation}

Several comments are in order here. 
If we apply
Hamilton's equations to the energy relation in Eq. (\ref{energy}),
i.e., $dz/dt=\partial E/\partial{p_z}$ and 
$dp_z/dt=-\partial E/\partial{z}$,
we get expressions
for $dz/dt$ and $d p_z/dt$ that agree with the expectation
values of the corresponding
evenised operators $[d\hat z/dt]$ and $[d \hat p_z/dt]$.
But the real number
$p_z$ appearing above in Eq. (\ref{energy}) need not be the canonical 
momentum.  It represents the expectation value of $[\hat p_z]$
whereas the canonical momentum would be associated with
the expectation value of $\hat p_z$.
Secondly, ignoring for now the distinction between $\hat p_z$ and 
$[\hat p_z]$, Ehrenfest's theorem for a relativistic system would imply
$d\langle\hat z\rangle /dt=\langle\partial \hat H/\partial{{\hat p_z}}\rangle$ 
and 
$d \langle \hat p_z\rangle /dt=-\langle\partial\hat H/\partial{\hat z}\rangle$,
and not 
$d\langle\hat z\rangle /dt=
\partial \langle\hat H\rangle /\partial\langle{\hat p_z}\rangle$ and 
$d \langle \hat p_z\rangle /dt=-\partial\langle\hat H\rangle/
\partial\langle{\hat z}\rangle$.  Nevertheless the energy relation
in Eq. (\ref{energy}) behaves as a quantum corrected classical Hamiltonian,
i.e., one may apply Hamilton's equations to it to obtain the velocity and
the force.
However,
in general, one can not presume that $E$ is the quantum corrected
classical Hamiltonian and indeed in Ref. \cite{br} 
with the unrotated Lagrangian one can not treat $E(p_z,z)$ as the
quantum corrected
classical Hamiltonian
with $p_z$ as the 
canonical momentum.

In the literature \cite{prok,cline} there has been discussion on identifying
the canonical and kinetic momentum of the particle. One may presume that since
$\hat p_z$ is identified with the $-i\hat\partial_z$ 
operator it is conjugate to 
$\hat z$ and represents the canonical momentum.  
Furthermore $p_z$ acts as a canonical momentum if the energy relation of
Eq. (\ref{energy}) is treated as a classical Hamiltonian.
Now our energy relation also agrees with 
with that obtained below using the Wigner formalism.
But that energy relation is necessarily in terms of the kinetic 
momentum as 
the Wigner function is always written in
terms of the kinetic momentum of the system (see Sec. 3 of Ref. \cite{vge}).
This indicates
that our real number $p_z=\langle [\hat p_z]\rangle$ 
in the energy relation above is the kinetic momentum of the system.
Thus the evenised canonical momentum  in this study
becomes
equivalent to the canonical momentum and the kinetic momentum, unlike
in Ref. \cite{br}.

The single particle trajectory has been obtained in a frame in which
the $x$ and $y$ components of the momentum are 0.
For a single particle such a frame may be defined but not so for a collection
of particles.  
Therefore the treatment above may be treated as that for a
1+1 dimensional universe only.
It is non-trivial to derive as above the kinetic equation 
for a 3+1 dimensional universe because of the presence of spin.

\section{Kinetic equation using the Wigner formalism}
We now present a more formal field theoretic derivation of the kinetic
equation in a manner analogous to that adopted in Ref. \cite{prok} for the
unrotated Lagrangian.
As stated in the Introduction,  one can obtain the kinetic equation 
from the equation of motion
for the Wigner transform of the Wightman function.
We shall closely follow the derivation in Ref. \cite{prok} for
the unrotated Lagrangian.  The Wightman function is defined as
\begin{equation}
G^<_{\alpha\beta}(u,v) \equiv
       i\left< \bar{\psi}_\beta(v)\psi_\alpha(u) \right>\,.
\end{equation}
Starting from the Dirac equation for the fermions in the electroweak plasma, 
one 
can obtain the equation of motion for the Wightman function by multiplying the 
Dirac equation by $i\bar\psi$, passing it through the differential
operator and taking the expectation value.
\begin{equation}
[i\dslash_u -{m(u)\over\hbar} +{\partial_\mu\theta(u)\over2}\gamma^\mu\gamma^5]_{\rho\alpha}
i \left < \bar{\psi}_\beta(v)\psi_\alpha(u) \right>=0
\end{equation}
Subsequently, performing a Wigner transform of $G^<(u,v)$, one obtains
$\cG^<(x,k)$.  
\begin{equation}
\cG^<(x,k) \equiv \int d^{\,4} r \, e^{(i/\hbar)k\cdot r} G^<(x + r/2,x-r/2),
\end{equation}
where $r=u-v$.
Unlike the Wightman function which is a function of two
spacetime variables, $\cG^<(x,k)$ is in a mixed representation, i.e., it is a function
of $x$ and $k$ and hence can finally give a kinetic equation for 
particle and antiparticle distribution functions which are functions of
position and momentum.  $\cG^<(x,k)$
satisfies 
\begin{equation}
\left(\hat{\kslash} - {\hat m}(x) +{\hbar\hat {\partial_\mu\theta}\over2} 
\gamma^\mu\gamma^5\right) \,
\cG^< = 0,
\label{GWigeom}
\end{equation}
where
\begin{eqnarray}
{\hat k}_\mu &\equiv& k_\mu + \frac{i\hbar}{2}\partial_\mu \\
{\hat m}&\equiv& m
   e^{-\frac{i\hbar}{2}\stackrel{\!\!\leftarrow}{\partial_x}\cdot\,\partial_k}\\
\hat {\partial_\mu\theta}  &\equiv&  (\partial_\mu\theta)
e^{-\frac{i\hbar}{2}\stackrel{\!\!\leftarrow}{\partial_x}\cdot\,\partial_k}
\end{eqnarray}
(Clearly, $\, \hat {} \,$ in this section 
is defined differently than in the previous
section.)
We now assume that the bubbles are large enough that we can take
the wall to be planar (and in the $x-y$ plane).
Then, working in a wall frame, as in Ref. \cite{prok}, in which the momentum
of the particles parallel to the wall vanishes,
$S^3i\gamma^0{\cal G}^<  =  si\gamma^0\cal G^<$.  Therefore, since
$(I-s S^3) i\gamma^0{\cal G}^<=0$,
$\cG^<(x,k)$ may be expressed, in the chiral representation, as  
\begin{eqnarray}
-i\gamma^0\cG^<& =& g_s^< \otimes\frac12(I+s\sigma^3) \\
& =&  \frac12 \left( g^{s}_0 I +  g^{s}_i \rho^i \right)
 \otimes\frac12(I+s\sigma^3)
 ,
\label{Wigdecomp}
\end{eqnarray}
where $\sigma^i\equiv\rho^i$ 
are the Pauli
matrices and $g_{0,i}^s$ are functions of $k^3,x^3,t$ and $s$.  
\footnote{Here we have used the definition of the direct product as in
Ref. \cite{prok2,IZ}.  The order of the two terms in the product is reversed
in Ref. \cite{prok}.  Also,
the above decomposition of $\cG^<(x,k)$ is as in Refs. \cite{prok,prok3d}
which differs by a factor of 2 from the decomposition in Ref. \cite{prok2}.}
As discussed earlier, the Wigner function above
is now for a 1+1 dimensional universe.

This decomposition allows one to reduce the 4x4 equation for the Wigner
function, Eq. (\ref{GWigeom}) to a 2x2 equation for $g_s^<$.  Expanding the
operator acting on $-i\gamma^0\cG^<$ as a direct product of 2x2 matrices
(see Ref. \cite{IZ}), we
get
\begin{equation}
[\hat k^0
+\frac {\hbar s} {2}
\hat {\partial_3\theta}
-s\hat k^3\rho^3-\hat m \rho^1 
]g_s^<=0 \, .
\label{gseqn}
\end{equation}
(Here we have used the definition of the gamma matrices as in 
Refs. \cite{prok,prok2}, which differs from the definition
of $\gamma^0$ and $\gamma^5$ in Ref. \cite{IZ} by a minus sign.  The metric
is (1,-1,-1,-1).) 
The corresponding equation obtained in Ref. \cite{prok} for the unrotated
Lagrangian is
\begin{equation}
[\hat k^0-s\hat k^3\rho^3-\hat m_0 \rho^1 -\hat m_5\rho^2
]g_s^<=0
\end{equation}
where
\begin{equation}
{\hat m_{0,5}}\equiv m_{R,I}
   e^{-\frac{i\hbar}{2}\stackrel{\!\!\leftarrow}{\partial_x}\cdot\,\partial_k}\,,
\end{equation}
where $m_R$ and $m_I$ are $|m|\cos\theta$ and $|m|\sin\theta$ respectively.
Expanding $g_s^<$ in terms of I and the Pauli matrices as in 
Eq. (\ref{Wigdecomp}) and multiplying Eq. (\ref{gseqn}) by I and $\rho^i$ and
taking the trace gives
\begin{eqnarray}
({\hat k}^0 +\frac {s\hbar}2 \hat{\partial_3\theta}) g^{s}_0  
- \hi s{\hat k}^3 g^{s}_3
            - \hi {\hat m} g^{s}_1  &=& 0
 \\
({\hat k}^0 +\frac {s\hbar}2 \hat{\partial_3\theta}) g^{s}_3  - \hi s{\hat k}^3 g^{s}_0
            - i   {\hat m} g^{s}_2  &=& 0
 \\
({\hat k}^0 +\frac {s\hbar}2 \hat{\partial_3\theta}) g^{s}_1  + i   s{\hat k}^3 g^{s}_2
            - \hi {\hat m} g^{s}_0  &=& 0
 \\
({\hat k}^0 +\frac {s\hbar}2 \hat{\partial_3\theta}) g^{s}_2  - i   s{\hat k}^3 g^{s}_1
            + i   {\hat m} g^{s}_3  &=& 0.
\end{eqnarray}
These equations are similar to Eqs. (2.16)-(2.19) of Ref. \cite{prok} but with
$\hat k^0\rightarrow \hat k^0+\frac {s\hbar}2 \hat{\partial_3\theta}$, $\hat m_0
\rightarrow \hat m$ and $\hat m_5\rightarrow 0$.
One can independently set the real and imaginary parts of these equations
to 0.  The imaginary parts contain time derivatives and so provide the
kinetic equations.  The real parts provide constraint equations.  
Assuming that $m$ and $\partial_3\theta$ vary slowly in the Higgs bubble
wall we may expand $\hat m$ and $\hat {\partial_3 \theta}$ to $O(\hbar)$ as
\begin{eqnarray}
\hat m&=&m+\frac {i\hbar}2 m'\partial_{k_z}\\
\hat {\partial_3\theta}&=&\theta^\prime+\frac {i\hbar}2 \theta^{\prime\prime}
\partial_{k_z} \,.
\end{eqnarray}
Henceforth, we define $z\equiv x^3$ and $k_z\equiv k^3$, and
$\prime$ denotes $\partial/\partial x^3=\partial/\partial z$
and $\partial_{k_z}=\partial/\partial k^3$.
$\stackrel{\!\!\leftarrow}{\partial_x}\cdot\,\partial_k=
-\stackrel{\!\!\leftarrow}{\partial_z}\cdot\,\partial_{k_z}$.
The constraint equations then are
\begin{eqnarray}
&& (k^0 + \frac {s\hbar}2\thetaprime) g^{s}_0 - sk_z g^{s}_3 - m g^{s}_1 
 = 0\label{constr1}
 \\
&& (k^0 + \frac {s\hbar}2\thetaprime)g^{s}_3 - sk_z g^{s}_0 
+ \frac\hbar2 m' \partial_{k_z} g^{s}_2
                           = 0
\label{constr2}\\
&& (k^0 + \frac {s\hbar}2\thetaprime)g^{s}_1 
+ \frac{s\hbar}{2} \partial_z g^{s}_2 - m g^{s}_0  = 0
\label{constr3} \\
&& (k^0 + \frac {s\hbar}2\thetaprime)g^{s}_2 
- \frac{s\hbar}{2} \partial_z g^{s}_1
                - \frac\hbar2 m' \partial_{k_z} g^{s}_3 = 0.
\label{constr4}\end{eqnarray}
The above equations imply relations between $g_0^s$ and 
$g_i^s$.  Using Eqs. (\ref{constr2},\ref{constr3},\ref{constr4}) iteratively to
$O(\hbar)$ we express
the $g_i^s$ in terms of $g_0^s$.  Substituting this in Eq. (\ref{constr1}) gives
\begin{equation}
\left( (k^0+\frac {\hbar s}2\theta')^2 -  k_z^2 - m^2  
       \right) g^{s}_0 = 0 \, ,
\label{g0eqn}
\,.
\end{equation}
This implies 
\begin{equation}
\Omega^2_s \; \equiv \;
 (k^0+\frac {\hbar s}2\theta')^2 -  k_z^2 - m^2  
 = \; 0,
\label{Omega}
\end{equation}
whose solutions are $k^0=\pm \omega_{s\pm}(z,k_z)$ with
\begin{equation}
     \omega_{s\pm}(z,k_z) = \omega_0 \mp  \frac {\hbar s}2{\theta'}\,,
     \qquad \qquad \omega_0 = \sqrt{ k_z^2 + m^2}
\label{dispersion1} 
\end{equation}
$\omega_{s+}(z,k_z)$ and $\omega_{(-s)-}(z,-k_z)$ will
represent the energy of particles/antiparticles
as they pass through the Higgs bubble wall.  This agrees
with Eq. (\ref{energy}).

The kinetic equation for $g_0^s$ is
\begin{equation}
\partial_t g_0^s +\frac {\hbar s}2 \theta^{\prime\prime}\partial_{k_z} g_0^s
+s\partial_z g_3^s -m'\partial_{k_z} g_1^s=0 \,.
\end{equation}
Using the constraint equations to re-express $g_i^s$ in terms of $g_0^s$, 
the kinetic equation for $g_0^s$ implies
\begin{equation}
         \partial_t  g^{s}_0 + k_z\partial_z \left({g^{s}_0\over
k^0+{\hbar s\thetaprime\over2}}\right)
-\frac{1}{2} {m^2}^{\,\prime} 
\partial_{k_z} \left(g^{s}_0\over k^0+{\hbar s\thetaprime\over2}\right) 
+\frac {\hbar s}2\thetadprime \partial_{k_z} g^{s}_0
= 0 \, .
\end{equation}
Analogous to Ref. \cite{prok3d}, Eq. (\ref{g0eqn}) implies that
$g_0^s$ may be written as  
\begin{equation}
g^{s}_0 = 4\pi N  |k^0|
\delta(\Omega_s^2)\, . 
\label{g0P}\end{equation}
However, for reasons discussed below, we define
\begin{eqnarray}
g^{s}_0 &=& 4\pi N  |k^0+\hbar s\thetaprime/2| 
\delta(\Omega_s^2)\\
  &=& \sum_\pm {4\pi }
                 \, N \frac{|k^0+\hbar s\thetaprime/2|}{2\omega_0}\,\delta(k^0 \mp \omega_{s\pm})\,.
\label{g01}
\end{eqnarray}
$N$ is a function of $k^0,k_z,z,t$ and spin $s$.  

One now substitutes for $g_0^s$ from Eq. (\ref{g01}) in the kinetic
equation and integrates over positive 
and negative energies separately to 
get 
\begin{equation}
        \partial_t f_{s\pm}
      + \left( \frac{k_z}{\omega_0} 
\right) 
\partial_z f_{s\pm}
      + \left(-\frac{m^{2\,\prime}}{2\omega_0} 
+\frac {\hbar s}2 \thetadprime
                   \right)
 \partial_{k_z} f_{s\pm} = 0.
\label{boltzmann2-sc}
\end{equation}
where $f_{s\pm}(k_z,z,t)$ are
functions given by
\begin{eqnarray}
     f_{s+} &\equiv& N(\omega_{s+},k_z,z,t)
\\
     f_{s-} &\equiv& 1 - N(-\omega_{(-s)-},-k_z,z,t)
\label{fs}
\end{eqnarray}
If we can identify $f_{s\pm}$ with the particle and antiparticle
distribution functions \footnote{See footnote [\ref{negspin}].},
the above equations have the form of the classical kinetic equation, with
quantum corrections in the coefficients of the derivatives of $f_{s\pm}$.

The identification of the particle and antiparticle 
distribution functions can be made in two ways.  
One can argue that the distribution function is that quantity
which gives the number density on integration over momenta.
Defining the current as
\begin{equation}
j^\mu(x)=\left< \bar\psi(x)\gamma^\mu\psi(x) \right>
\end{equation}
we rewrite it as in Sec. 4.1 of Ref. \cite{0406140} as 
\begin{equation}
j^\mu(x)
  = -\Tr [\gamma^\mu i G^<(x,x)]
  = - \int \frac{d^2k}{(2\pi)^2} \Tr \gamma^\mu i\cG^<(k,x)
\end{equation}
Substituting for the Wigner function from Eq. (\ref{Wigdecomp}),
and using $\Tr A\otimes B=\Tr A \, \Tr B$ and the tracelessness of the
Pauli matrices, we get
\begin{equation}
j^0 = 
 \sum_{s=\pm 1} 
\int \frac{d^{\,2}k}{(2\pi)^2}
             g^s_0(k^0,k_z,z,t)\,.
\end{equation}
If we use the decomposition in Eq. (\ref{g01}) for $g^s_0$ we find
that we can write $j^0$ as $j^0_{+} - j^0_{-}$ where
\begin{equation}
j^0_{\pm}=\sum_{s=\pm1}j^0_{s\pm} =
\sum_{s=\pm1}
\int \frac{dk_z}{(2\pi)}
        f_{s\pm}     \,.
\label{j0}
\end{equation}
(We have ignored the vacuum contribution to $j^0$.)
Thus if we 
define particle and antiparticle number densities (per unit length)
to be $j^0_\pm$
it seems appropriate to identify
$f_{s\pm}$ with particle and antiparticle
distribution functions.  
Alternatively, 
one can argue that the 
quantum corrected kinetic equation should have the same form as the classical
kinetic equation, i.e., it should contain only derivatives of the distribution
function and not terms proportional to the distribution function.  
Since the equation for $f_{s\pm}$ satisfies this criterion it again seems
correct to identify $f_{s\pm}$ with particle and antiparticle distribution 
functions.  
However, note that if we had used Eq. (\ref{g0P}) as the decomposition for
$g_0^s$ neither of the above criteria would be satisfied.  This is why
we chose the decomposition in Eq. (\ref{g01}).
We further note that with the $f_{s\pm}$ identified as particle and antiparticle
distribution functions Eq. (\ref{boltzmann2-sc}) 
above agrees with the kinetic equation
obtained in Sec. II.

We briefly comment on the application of the above ideas to the exposition
in Refs. \cite{prok,prok2,0406140} for the unrotated Lagrangian.  
While $f_{s+}$ in these
works satisfies the second criterion above, i.e., the quantum corrected kinetic
equation has the form of the classical kinetic equation, the quantity whose
integral over momentum gives $j^0_+$ are $f_{s+}/Z_{s+}$, i.e.,
\begin{equation}
j^0_{s+}=\int {dk_z\over 2\pi} f_{s+}/Z_{s+} \,,
\label{j01}\end{equation}
where 
\begin{equation}
Z_{s+} \equiv \frac{1}{2 \omega_{s+}\,} |\partial_{k^0}
                      \Omega^2_s|_{k^0=\omega_{s+}} \,.
\end{equation}
(See
Eq. (2.41) of Ref. \cite{prok} but with the normalisation of $g_{0}^s$ as in
Ref. \cite{prok3d}.  
\footnote{The $Z$ factor is absent in Eq. (4.10) of Ref. \cite{0406140}
because the quantity $\delta f^v_s$ in the numerator is already $O(\hbar)$.
})
If one rewrites the kinetic equation in terms of the variable 
$f_{s+}/Z_{s+}$ one gets an additional term in the kinetic equation.

But let us now consider
the quantity which gives the number density of particles
on integration over 4-momentum, or rather, 2-momentum
in our 1+1 dimensional case, after
multiplying with kinetic energy and a delta function containing the mass shell
condition, i.e.,
\begin{equation}
j^0_{s+}=\int {dk_z \, dk^0 \over (2\pi)^2} \,2K^0 \,2\pi \delta(\Omega^2)\, 
F_s(x,k) \,,
\label{j02}
\end{equation}
where $K^0$ is the kinetic energy.
When converting
$\delta(\Omega^2)$ to $\delta(k^0-\omega_{s+})$
one ultimately gets a factor
of $Z_{s+}$ in the denominator.
(For the unrotated Lagrangian,
the kinetic energy $K^0$ equals the total energy $\omega_{s+}$, as evidenced by
the definition of the kinetic momentum in Eq. (2.14) of Ref. \cite{cline} as
$\omega v_g$, where $\omega$ is the total energy and $v_g$ is the velocity.)
Now if 
we choose to define the particle distribtuion functions
as $F_s(x,k)$ with $k^0$ replaced by $\omega_{s+}$, i.e., in terms of
the function in the integrand of $j^{0}_{s+}$ when expressed as an integral
over $k_z$ and $k^0$ as in Eq. (\ref{j02})
rather than as an integral over only $k_z$ as in Eq. (\ref{j01}),
then we can identify
$f_{s+}$, rather than $f_{s+}/Z_{s+}$, with the particle distribution
function and one also has a `standard' form for the kinetic equation
for this particle distribution function.  
Therefore 
this prompts us to modify our first scheme of identification
of the particle distribution function.  It does not affect the case of the
rotated Lagrangian.

The transport equations relevant for electroweak baryogenesis 
for the unrotated Lagrangian are obtained
in Eqs. (5.2), and (5.8) and (5.9) of Ref. \cite{prok}
from the kinetic equations for $g_0^s$ and
$g_3^s$, which agree with taking the zeroth and first
moments of the kinetic equation
for $f_{s+}$, i.e., by integrating the kinetic equation over $k_z$ after
multiplying by 1 and the velocity $v$.  For the rotated
Lagrangian considered in this article we obtain the kinetic equation using
both methods and they agree (for us, $v\equiv k_z/\omega_{0}$).  The equations
we obtain are
\begin{eqnarray}
   \partial_t n_{s\pm} + \partial_z(n_{s\pm} u_{s\pm}) &=& 0\\
   \partial_t \left(n_{s\pm} u_{s\pm}  \right)
     + \partial_z \left(n_{s\pm} \langle v_{s\pm}^2 \rangle \right)
+\half m^{2\prime} {\cal I}_{2s\pm} 
- \half s(m^2\thetaprime)' {\cal I}_{3s\pm} &=&0
\end{eqnarray}
where
\begin{eqnarray}
n_{s\pm}&=& \int {dk_z\over 2\pi} f_{s\pm}\\
n_{s+} \langle v_{s+}^p \rangle &\equiv&
        \int{dk_z\over 2\pi}
       \left( \frac{k_z}{\omega_0} \right)^p f_{s\pm} \,,
\end{eqnarray}
and $u_{s\pm}\equiv \langle v_{s\pm} \rangle$,
and 
\begin{equation}
{\cal I}_{ps\pm} = 
                  \int \frac{dk_z}{2\pi }
                   \frac{f_{s\pm}}{\omega_{(\pm s)\pm}^p}
\quad (p=2,3)\, .
\end{equation}
It is interesting to note that the form of the above transport equations for
the rotated Lagrangian are identical
to those in Ref. \cite{prok} for the unrotated Lagrangian
(if one amends Ref. \cite{prok} to identify antiparticles of spin $s$
with negative energy particles of spin $-s$).

\section{Conclusion}

In conclusion, we have obtained 
the energy relation and the kinetic equation 
for fermions and antifermions interacting with the
Higgs bubble wall during the electroweak phase transition using the method
of evenisation and the Wigner formalism.  Our results for the velocity and force
on the particle/antiparticle as they pass through the Higgs wall are the same
using both methods.  This indicates that evenisation is indeed a reliable method
to investigate the quantum corrections to the velocity and force acting
on particles and antiparticles as they traverse the bubble wall during
the electroweak phase transition.

\begin{acknowledgements}
We would like to thank Tomislav Prokopec and Steffen Weinstock 
for many useful
clarifications about their work.
\end{acknowledgements}

\begin{appendix}
{\bf \centerline {Appendix}}

We present below the derivations of the various evenised expressions
quoted in the text.  Standard formulae that we shall use are
$\hat\beta^2=-(\hat\gamma^0\hat\gamma^5)^2=(\hat\alpha^3)^2=1$.
The symbols square brackets are used below for both commutators and evenised operators
but the difference is clear from the context.

{\bf Evenised  
$\hat\beta$, $\hatalpha3$ and $\hat\gamma^0\hat\gamma^5$ to $O(\hbar^0)$:}

To $O(\hbar^0)$,
\begin{eqnarray}
2[\hat\beta]&=&\hat\beta + \hat\Lambda_0 \hat\beta \hat\Lambda_0\cr
            &=&\hat\beta + {\hat\alpha^3 p^3 +\hat\beta  \hat m 
\over E_0}
	    \hat\beta \hat\Lambda_0\cr
	    &=&\hat\beta + \hat\beta{-\hat\alpha^3 p^3 +\hat\beta  \hat m
\over E_0}
	     \hat\Lambda_0\cr
	     &=&\hat\beta + \hat\beta{
	    -\hat\Lambda_0 E_0
+2\hat\beta  \hat m\over E_0
}
	     \hat\Lambda_0\cr
	     &=&{2 \hat m\over E_0}\hat\Lambda_0 \, ,
\end{eqnarray}
where we have used $\hat\Lambda_0^2= 1+O(\hbar)$ in the last 
equality.  Since the lhs is even and the sign operator on the rhs is even,
$m(\hat z)$ may be replaced by $[m(\hat z)]$, which, as explained
in the text, is equal to $m([\hat z])$ to $O(\hbar)$. Therefore
\begin{equation}
[\hat\beta]={ m([\hat z]) \over E_0}\hat\Lambda_0 +O(\hbar)\, .
\label{evenbeta0}
\end{equation}

Similarly, to $O(\hbar^0)$,
\begin{eqnarray}
2[\hat\alpha^3]&=&\hat\alpha^3 + \hat\Lambda_0 \hat\alpha^3 \hat\Lambda_0\cr
            &=&\hat\alpha^3 + {\hat\alpha^3 p^3 +\hat\beta  \hat m 
\over E_0}
	    \hat\alpha^3 \hat\Lambda_0\cr
	    &=&\hat\alpha^3 + \hat\alpha^3{\hat\alpha^3 p^3 -\hat\beta  \hat m
\over E_0}
	     \hat\Lambda_0\cr
	     &=&\hat\alpha^3 + \hat\alpha^3{
	    -\hat\Lambda_0 E_0
+2\hat\alpha^3 p^3\over E_0
}
	     \hat\Lambda_0\cr
	     &=&{2 \hat p^3\over E_0}\hat\Lambda_0 \, 
	     ={2 \hat p_z\over E_0}\hat\Lambda_0 \, ,
\end{eqnarray}
Again, since the lhs is even and the sign operator on the rhs is even,
$\hat p_z$ may be replaced by $[\hat p_z]$. Therefore
\begin{equation}
[\hat\alpha^3]={ [\hat p_z]\over E_0}\hat\Lambda_0 +O(\hbar)\, .
\label{evenalpha0}\end{equation}

Analogously,
\begin{eqnarray}
2[\hat\gamma^0\hat\gamma^5]&=&\hat\gamma^0\hat\gamma^5 +
\hat\Lambda_0 \hat\gamma^0\hat\gamma^5 \hat\Lambda_0\cr
&=& \hat\gamma^0\hat\gamma^5 - \hat\gamma^0\hat\gamma^5
\hat\Lambda_0 \hat\Lambda_0\cr
&=&0\, .
\end{eqnarray}
Therefore,
\begin{equation}
[\hat\gamma^0\hat\gamma^5]=O(\hbar)\, .
\end{equation}

{\bf Evenised expressions for the velocity and the force
to $O(\hbar)$:}  

Since we wish to work till $O(\hbar)$ 
we now use the sign operator $\hat\Lambda$ defined as
\begin{equation}
\hat\Lambda={\hat H \over E}
\end{equation}
with $E$ defined to $O(\hbar)$ as in Eq. (\ref{energy}).  Now
$d\hat z/dt=-(i/\hbar)[\hat z,\hat H]=\hat \alpha^3$.  Then
\begin{eqnarray}
2[\hat\alpha^3]&=&\hat\alpha^3 + \hat\Lambda \hat\alpha^3 \hat\Lambda\cr
&=&\hat\alpha^3 + \hat\alpha^3 {
-\hat\Lambda E+
2\hat\alpha^3 \hat p^3 -\hbar\hat\thetaprime\hatS3
\over E}
\hat \Lambda\cr
&=&{2 \hat p^3 -\hbar\hat\thetaprime\hatalpha3\hatS3\over E} \hat\Lambda \, =
\,{2 \hat p_z \over E} \hat\Lambda 
-{\hbar\hat\thetaprime\hatalpha3\hatS3\over E}\hat \Lambda \,.
\end{eqnarray}
Since the lhs of the above equation
and the sign operator on the rhs are even we may replace
$\hat p_z$ by $[\hat p_z]$.  Since $\hatS3$ is also even,
we may similarly replace $\hat\thetaprime\hatalpha3$ by 
$[\hat\thetaprime\hatalpha3]$.
Since $\{\hattheta^\prime\}$ is proportional to $[\hattheta^\prime,\hat\Lambda]
\sim \hbar$ we may rewrite $[\hat\thetaprime\hatalpha3]$ as
$[\hat\thetaprime][\hatalpha3]$.  As discussed in the text for
$m(\hat z)$, $[\hat\thetaprime]=\thetaprime([\hat z]) +O(\hbar^2)$.  Further
we may use the expression for $[\hatalpha3]$ obtained above
in Eq. (\ref{evenalpha0}) upto $O(\hbar^0)$ in the rhs of the above 
equation.
Therefore we get, to $O(\hbar)$,
\begin{equation}
[d\hat z/dt]=\left({ [\hat p_z] \over E} -{\hbar\hatthetaprime[\hat p_z]
\hatS3\hat\Lambda_0\over2E_0^2}
\right )\hat\Lambda \, .
\end{equation}

The force $d \hat p_z/dt=-(i/\hbar)[\hat p_z,\hat H]=
-\hat \beta \hat m' +\hbar\hat\theta^\doubleprime\hatS3/2$.
Then
\begin{eqnarray}
2[\hat \beta  \hat m']&=&\hat\beta \hat m' 
+ \hat\Lambda \hat\beta\hat m' \hat\Lambda\cr
&=&\hat\beta \hat m' + \hat\beta \hat m' {
-\hat\Lambda E
+2\hat\beta  \hat m-\hbar\hattheta^\prime\hatS3
\over E}
\hat\Lambda  -\hat\beta\hat\alpha^3{[\hat p_z, \hat m']\over E} \, \hat\Lambda \cr
&=&{2  \hat m'  \hat m -\hbar\hatbeta  \hat m'\hattheta^\prime\hatS3
+i \hbar\hat\beta\hat\alpha^3  \hat m^{\prime\prime} \over E} \hat\Lambda \, .
\end{eqnarray}
Again, we may replace $ \hat m'  \hat m$ by $[ \hat m'  \hat m]$ as
both sides of the equation should be even.
As $\{ \hat m'\}\{ \hat m\}$ is $O(\hbar^2)$, Eq. (\ref{prod}) implies that
this reduces to
$[ m'(\hat z) ][ m(\hat z) ]$.  This 
may further be rewritten as
$ m'([\hat z]) m([\hat z]) $.
For the second term in the numerator, we replace
$\hatbeta  \hat m'\hattheta^\prime\hatS3$ by
$[\hatbeta  \hat m'\hattheta^\prime\hatS3]=
[\hatbeta  \hat m'\hattheta^\prime]\hatS3$.  This may be rewritten as
$[\hatbeta  \hat m'][\hattheta^\prime]\hatS3$.
As above, we replace $[\theta^\prime(\hat z)]$ by
$\theta^\prime([\hat z])$.  Since, by Eq. (\ref{oddA}),
$\{ \hat m'\}$ is $O(\hbar)$, $[\hatbeta  \hat m']$ can be rewritten
as $[\hatbeta][  \hat m']$, and $[\hatbeta]$ may be replaced by the
$O(\hbar^0)$ expression in Eq. (\ref{evenbeta0}) and $[  \hat m']$
by $ \hat m'([\hat z]) $.
For the third term, we 
replace $\hat\beta\hat\alpha^3  \hat m^{\prime\prime} $ by
$[\hat\beta\hat\alpha^3  \hat m^{\prime\prime}]$.  Since this term is
already of $O(\hbar)$ \,
we can write it as 
$[\hat\beta\hat\alpha^3][m(\hat z) ^\doubleprime]$ and further as
$[\hat\beta\hat\alpha^3] m^{\prime\prime}([\hat z])$.  
Now $\hat\beta\hat\alpha^3=\hat S^3
\hat\gamma^0\hat\gamma^5$ and since $\{ \hat S^3\}=0$, $[\hat\beta\hat\alpha^3]=
\hat S^3 [\hat\gamma^0\hat\gamma^5]=O(\hbar)$.  Therefore, this term is of
$O(\hbar^2)$ and we may ignore it.
Thus
\begin{equation}
[\hat \beta  \hat m']=\left( {m([\hat z]) ^{2\prime}\over 2E }
-{\hbar m([\hat z]) ^{2\prime}\theta'([\hat z])\hat\Lambda_0\hat S^3  
\over 4E_0^2}\right)\hat\Lambda \, .
\end{equation}

Therefore
\begin{eqnarray}
[{d\hat p_z/dt}]&=&
\left( -{ \hat m^{2\prime}\over 2E }
+{\hbar \hat m^{2\prime}\hat\theta'\hat\Lambda_0\hat S^3  
\over 4E_0^2}\right)\hat\Lambda 
+{\hbar\hat\theta^\doubleprime\hatS3\over2}\\
&=&
\left(- { \hat m^{2\prime}\over 2E }
+{\hbar \hat m^{2\prime}\hat\theta'\hat\Lambda_0\hat S^3  
\over 4E_0^2}
+{\hbar\hat\theta^\doubleprime\hatS3\over2}\hat\Lambda_0
\right)\hat\Lambda\, . 
\end{eqnarray}
$ \hat m$ and $\hat\theta$ above are functions of $[\hat z]$.

\end{appendix}


\end{document}